\documentclass{jetpl}
\twocolumn

\lat

\usepackage{graphicx}
\usepackage{bm}
\usepackage{color}

\renewcommand{\vec}[1]{{\bf #1}}

\def\ga{{\ \lower-1.2pt\vbox{\hbox{\rlap{$>$}\lower5pt\vbox{\hbox{$\sim$}}}}\ }}
\def\la{{\ \lower-1.2pt\vbox{\hbox{\rlap{$<$}\lower5pt\vbox{\hbox{$\sim$}}}}\ }}

\def\eps{\varepsilon}
\def\ua{\uparrow}
\def\da{\downarrow}

\def\beq{\begin{equation}}
\def\eeq{\end{equation}}

\input{psfig.sty}

\title{Triplet $p$-wave superconductivity in low-density extended Hubbard model with Coulomb repulsion}

\rtitle{Triplet $p$-wave superconductivity in low-density extended Hubbard model with Coulomb repulsion}

\sodtitle{Triplet $p$-wave superconductivity in low-density extended Hubbard model with Coulomb repulsion}

\author{M.Yu. Kagan$^1$, D.V. Efremov$^2$, M.S. Marienko$^3$, and V.V. Val'kov$^4$}

\rauthor{M.Yu. Kagan, D.V. Efremov, M.S. Marienko, and V.V. Val'kov}

\sodauthor{M.Yu. Kagan, D.V. Efremov, M.S. Marienko, and V.V. Val'kov}

\address{
  $^1$ P.L. Kapitza Institute for Physical Problems, Russian Academy of Sciences, Moscow 119334, Russia\\
  $^2$ Max-Planck-Institut f\"{u}r Festk\"{o}rperforschung, Stuttgart D-70569, Germany\\
  $^3$ Department of Physics and Astronomy, Hofstra University, Hempstead, New York 11549,USA\\
  $^4$ Kirenskii Institute of Physics, Krasnoyarsk, Russia
     }

\dates{\today}{\today}

\abstract{ We analyze superconducting instabilities in 3D and 2D
extended Hubbard model with Coulomb repulsion between electrons
on neighboring sites in the limit of low electron density
($n_{el} \rightarrow 0$) on simple cubic (square) lattice. We
show that in a realistic strong-coupling case  $U\gg V\gg W$
($U$ and $V$ are the onsite and the intersite Coulomb repulsions,
$W$ the bandwidth) the main SC instability corresponds to the
$p$-wave pairing and in the leading order is correctly described
by the equations obtained earlier in the absence of the intersite
Coulomb interaction $V=0$. }


\begin{document}

\maketitle
\section{Introduction}

One of the main challenges of the modern condensed matter physics
is to identify the origin of superconductivity in superfluid
$^{3}$He, heavy fermion compounds and Sr$_2$RuO$_4$, semimetals
and superlatices. A lot of the experimental data as well as
theoretical calculations suggest that the pairing results from
the electron-electron interaction. In this scenario, a Coulomb
repulsion is inverted into attraction due to the fermion
background and retardation effects. This was first suggested by
Kohn and Luttinger \cite{Kohn65} for a 3D system with point-like
repulsion. The authors of Refs. \cite{Fay68, Kagan88, Baranov92}
extended the analysis to 2D systems and took into account the
effects of long range Coulomb interaction in dense electron
plasma. Recently, the question about the role of full Coulomb
interaction for non-phonon mechanisms of superconductivity was
raised in connection with the HTSC physics by Alexandrov and
Kabanov \cite{Alexandrov11}, and it still demands very thorough
investigations both in the jellium and lattice models.

\begin{figure}[t]
     \centerline{   \includegraphics[width=6 cm]{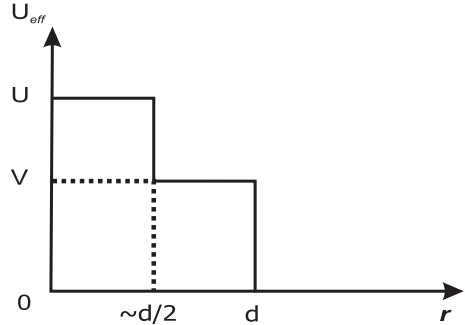}}
    \caption{Fig. 1. Effective interaction in the extended Hubbard model with Coulomb interaction on neighboring sites.}
    \label{fig:_interaction}
\end{figure}

In the present paper we consider the simplest and the most
repulsive (thus the most unfavorable for effective attraction and
SC) lattice model with the strong on-site Hubbard repulsion $U$
and the relatively strong additional Coulomb repulsion $V$ on the
neighboring sites (Fig. \ref{fig:_interaction}). We show that in
this model the $p$-wave superconductivity exists in both the 3D
and 2D case \cite{Fay68, Kagan88, Baranov92}. We assume the
following estimates: $U \sim e^2/\eps a_B$ for Hubbard $U$ and $V
\sim e^2/\eps d$ for Coulomb $V$. Here $a_B \sim \eps/me^2$ is
the Bohr radius, $\eps$ the effective dielectric permittivity,
$d$ the intersite distance. We assume that for $\eps \sim 1$:
$a_B \sim 0.5$ \AA, and $d \sim 3 - 4$ \AA.

In the simple 3D cubic lattice the bandwidth is $W=12t$ where $t$
is the hopping integral, and the electron mass at low density
(practically, empty lattice) $m=1/2td^2$. The uncorrelated
electron spectrum $\eps(p)=-2t(\cos p_x d+\cos p_y d+\cos p_z d)$
approximately has quadratic form $\eps(p)=-W/2 + p^2/2m$.
Similarly, the chemical potential measured from the bottom of the
band reads $\mu = -W/2 + \eps_F$, where $\eps_F=p_F^2/2m$ is the
Fermi energy, $p_F$ the Fermi momentum. If, as usual, we assume
$a_B \ll d$ (which, rigorously speaking, is valid at moderate
values of $\eps \ge 1$), then comparing the estimates for $U \sim
e^2/\eps a_B$, $V \sim e^2/\eps d$, and $W \sim 1/md^2$ in the
limit $a_B/d \ll 1$ we come to the following hierarchy of
parameters: \beq\label{parameter_range} U\gg V \gg W. \eeq Note
that some important SC systems possibly including HTSC could have
large values of $\eps$ and thus be in difficult intermediate
regime.

In this paper we construct the theory for the SC instability in
the parameter range (\ref{parameter_range}) and at low electron
density $n_{el}\rightarrow 0$ (or subsequent gas parameter $p_F d
\ll 1$), neglecting an important question of the microscopic
phase separation of Mott-Hubbard type \cite{Hubbard63, Nagaev67}
with FM polarons inside the AFM-matrix, and that of Verwey
type \cite{Verwey39, Kagan01} with metallic polarons  inside the
charge-ordered matrix. These instabilities towards nanoscale
phase separation arise in the model under condition
(\ref{parameter_range}) close to $n_{el}\rightarrow 1$ for the
Mott-Hubbard and $n_{el}\rightarrow 1/2$ for the Verwey type of
phase separation.

In the following we show that the leading SC instability at
$n_{el}\rightarrow 0$ corresponds to the triplet $p$-wave pairing
and in the leading order of the gas parameter \cite{Galitskii58}
$p_F d$ is described by the expressions obtained in Refs.
\cite{Fay68, Kagan88, Baranov92} for the low density Hubbard
model in the absence of Coulomb interaction (at  $V=0$). We
review the 2D case and present analogous results for the $p$-wave
pairing in the strong coupling case \cite{Chubukov93, Efremov00-1}
which is also in accordance with the low-density Hubbard model in
the absence of $V$  (at $V=0$).

\section{The model}

We consider the Hamiltonian
\begin{eqnarray}\label{Hamiltonian}
\hat{H}'&=&\hat{H}-\mu \hat{N}=-t\sum_{<ij>\sigma} c^\dagger_{i\sigma} c_{j\sigma} + U \sum_i n_{i\ua} n_{i\da} \nonumber\\
 &+&\frac{V}{2} \sum_{<ij>} n_i n_j -\mu \sum_{i\sigma} n_{i\sigma},
\end{eqnarray}
where $n_{i\sigma}=c^\dagger_{i\sigma} c_{i\sigma}$ is the
electron  density on site $i$ with spin projection $\sigma$.
After Fourier transformation, the Hamiltonian reads:
\begin{eqnarray}\label{Hamiltonian_Fourier_transformed}
\hat{H}'&=&\sum_{\vec{p}\sigma} [\eps (p) -\mu] c^\dagger_{\vec{p}\sigma} c_{\vec{p}\sigma} + U \sum_{\vec{p}\vec{p}'\vec{q}} c^\dagger_{\vec{p}\ua}c^\dagger_{\vec{p}'+\vec{q}\da}c_{\vec{p}+\vec{q}\da}c_{\vec{p}'\ua} \nonumber\\
&+&\sum_{\vec{p}\vec{p}'\vec{q}\sigma\sigma'} V(\vec{p},\vec{p}')
c^\dagger_{\vec{p}\sigma}c^\dagger_{\vec{p}'+\vec{q}\sigma
'}c_{\vec{p}+\vec{q}\sigma '}c_{\vec{p}'\sigma},
\end{eqnarray}
where \beq\label{Vpp} V(\vec{p},\vec{p}') = V [ \cos(p_x - p_x
')d + \cos(p_y - p_y ')d + \cos(p_z - p_z ')d ]. \eeq In analogy
with Ref. \cite{Kagan94} it is useful to expand the effective
interaction $U_{eff} = U + 2 V (\vec{p},\vec{p}')$ into the sum
of the $s$-wave and $p$-wave partial harmonics.

At the low density $pd\ll 1$ the expansion up to quadratic terms
gives effective interactions for $s$-wave and $p$-wave harmonics
correspondingly: \beq\label{Ueff_s_p} U_{eff}^s = U + 6V + o(p^2
d^2), {\mbox{~and~~}} U_{eff}^p =  2 V  \vec{p}\vec{p}'d^2. \eeq
In the
strong-coupling case $U\gg V \gg W$  it is convenient to
renormalize $U_{eff}^s$ and $U_{eff}^p$  in terms of vacuum
Kanamori $T$-matrices $T_s$ and $T_p$ \cite{Kanamori63}. To do
that we solve the Bethe-Salpeter equation in
vacuum \cite{Landau77}. This yields \cite{Baranov92} in the
low-energy sector: \beq\label{Ts} T_s =
\frac{(U+6V)d^3}{1+(U+6V)d^3\int\frac{d^3 p}{(2
\pi)^3}\frac{1}{2\eps_p}}\sim \frac{(U+6V)d^3}{(1+\beta_s)}, \eeq
where $\beta_s \sim \frac{(U+6V)}{8\pi t}>0$ is the Born
parameter for the $s$-wave channel, and we neglect the antibound
state which corresponds to the pole of the $T$-matrix at high
energies $E\sim U$ \cite{Hubbard63, Anderson90}.

We can introduce the $s$-wave scattering length
\beq\label{as}
a_s = \frac{m T_s}{4 \pi} = \frac{T_s}{8\pi t d^2} \sim \frac{\beta_s d}{(1+\beta_s)},
\eeq
and in the strong-coupling limit $\beta_s \gg 1$, evidently, $a_s \sim d$ (see Ref. \cite{Baranov92}).

Correspondingly, the 3D gas parameter of Galitskii \cite{Galitskii58}:
\beq\label{lambdas}
\lambda_s = \frac{  2  a_s p_F}{\pi} \approx \frac{2p_F d}{\pi}.
\eeq
Note that the same result for the $s$-wave scattering length is valid in the strong-coupling
low-density limit of the Hubbard model without Coulomb interaction $V=0$.

Similarly, for the $T$-matrix in the $p$-wave channel
\beq T_p =
 2 A_p \vec{p}\vec{p}'d^2
 \eeq
we get \beq A_p = \frac{Vd^3}{1+ Vd^3 \int \frac{d^3 p}{(2\pi)^3}
\frac{p^2 d^2}{3} \frac{1}{2\eps_p}}=\frac{Vd^3}{(1+\beta_p)},
\eeq
where \beq \beta_p = \eta\frac{V}{W}>0 \eeq is the
dimensionless Born parameter for the $p$-wave channel and $\eta
\sim 1$ is a numerical coefficient.

Introducing the $p$-wave scattering length \beq a_p =
\frac{A_p}{8\pi t d^2} = \frac{Vd}{8\pi t (1+\beta_p)} \eeq we
obtain in the strong coupling case $\beta_p \gg 1$ \beq a_p \sim
d. \eeq Thus \beq \frac{mT_p}{4\pi}= 2 a_p \vec{p}\vec{p}'d^2
\sim  d^3 \vec{p}\vec{p}' \sim  d^3 pp' \cos\theta , \eeq
where
$\theta = \widehat{\vec{p}\vec{p}'}$, and thus the dimensionless
$p$-wave gas parameter reads: \beq\label{lambdap} \lambda_p \sim
p_F d^3 p_F^2  \sim (p_F d)^3. \eeq Note that the estimate
(\ref{lambdap}) is natural for the $p$-wave harmonics of the
scattering amplitude for slow ($p_F d <1$) particles in vacuum
\cite{Landau77}.

\section{Bethe-Salpeter integral equation for $T_c$}
According to Landau-Thouless criterion for SC \cite{Lifshitz80},
\beq\label{Gamma} \Gamma_l =
\frac{\tilde{\Gamma}_l}{1+\tilde{\Gamma}_l \ln \left(\frac{2
e^C\eps_F}{ \pi T_c}\right)}, \eeq where $C\approx 0.58$ is the
Euler constant, $\Gamma$ the total vertex for the Cooper channel,
$\tilde{\Gamma}$ the irreducible bare vertex, and $l$ the orbital
moment of the Cooper pair.

\begin{figure}[t]
      \centerline{\includegraphics[width=8 cm]{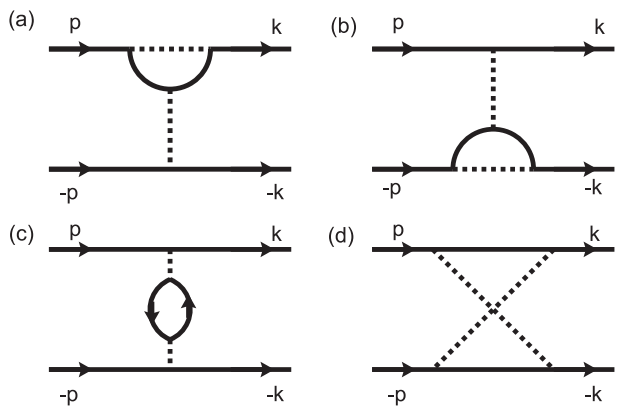}}
     \caption{Fig. 2. Irreducible diagrams in the second order of the $s$-wave gas parameter $\lambda_s$ which are nonzero only in the presence of fermion background (at $\eps_F \neq 0$).}
    \label{fig: 2_order}
\end{figure}

The critical temperature $T_c$ is given by the pole of
(\ref{Gamma}). If $\tilde{\Gamma}<0$ for several values of $l$,
then the actual symmetry of the superconducting state corresponds
to the highest $T_c$. According to Kohn and Luttinger
\cite{Kohn65}, in the absence of the Coulomb interaction ({\it
{i.e.}} of $\lambda_p$ (\ref{lambdap})) $\tilde{\Gamma}_{l\neq
0}$ is given by the sum of four irreducible diagrams (see Fig.
\ref{fig: 2_order}) which are of the second order of the $s$-wave
gas parameter $\lambda_s$.

At the same time, for $l=1$ due to the presence of Coulomb
repulsion $V$:  \beq\label{Gamma1} \tilde{\Gamma}_{l=1} =
\lambda_s^2 \Pi_{l=1}^{d}  +\lambda_p, \eeq where $\Pi^d = \Pi
(\vec{p}+\vec{k})$ is the exchange diagram (see Fig. \ref{fig:
2_order} d.), $\lambda_p$ is the bare vertex due to the $p$-wave
vacuum contribution of the intersite Coulomb interaction $V$.

As shown in Ref. \cite{Kohn65} for contact interaction
$\lambda_s$ the first three diagrams in the Fig. \ref{fig:
2_order} exactly cancel each other, and the resulting
$\tilde{\Gamma}_{l=1}$ is given by the fourth, exchange diagram
(see (\ref{Gamma1})).

\begin{figure}[t]
       \centerline{\includegraphics[width=9 cm]{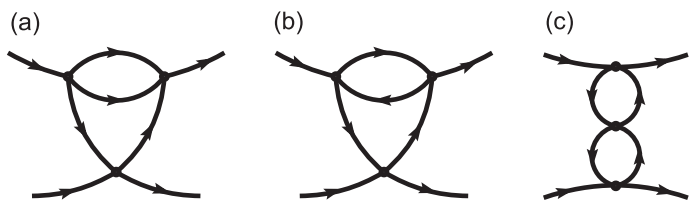}}
    \caption{Fig. 3. Irreducible diagrams in the third order of gas parameter for the Cooper channel.}
    \label{fig: 3_order}
\end{figure}

An exact evaluation of simple integrals shows \cite{Fay68, Kagan88, Baranov92} that for the exchange diagram
 $\lambda_s^2 \Pi_{l=1}^{d}  = -\frac{\lambda_s^2}{13}<0$
 which corresponds to the attraction and cannot be overcompensated by the repulsive bare vertex contribution $\lambda_p \sim \lambda_s^3$. This contribution only changes the next term in the expansion of $\tilde{\Gamma}_{l=1}$ in terms of gas parameter and, in fact, is the corrections to main exponent. To be specific (see Ref. \cite{Efremov00-1,Efremov00-2} and Fig. \ref{fig: 3_order}):
\beq\label{Gamma1_expanded} \tilde{\Gamma}_{l=1} =
-\frac{\lambda_s^2}{13} - \left( \frac{\lambda_s^3}{3} -
\lambda_p \right) + o(\lambda_s^4). \eeq

Let us repeat again that the result (\ref{Gamma1_expanded}) with
the bare vertex $\lambda_p \sim \lambda_s^3 \sim (p_F d)^3$ is to
some extent evident because according to quantum mechanics
\cite{Landau77} for slow particles in vacuum the $p$-wave
harmonic of the scattering amplitude is of the order
$(ap_F)^{2l+1} \sim (ap_F)^3$ at $l=1$ and $p\sim p_F$. Thus the
repulsive term in our case does not overcompensate the
Kohn-Luttinger attractive contribution which arises only in
fermion substance (when $\eps_F \neq 0$) and is proportional to
$(ap_F)^2$. The only peculiarity of the lattice is that $a \sim
d$ at the large Hubbard $U \gg W$ and at low density of
electrons, and hence $\lambda_s^2 \sim (p_F d)^2 \ll 1$.

Thus even at the most repulsive (and thus unfavorable for
effective attraction and SC) hierarchy of parameters $U \gg V \gg
W$ the presence of the Coulomb repulsion $V$ does not change the
main exponent for the $p$-wave critical temperature which reads
$T_{c1}\sim \eps_F \exp \left(-\frac{13}{\lambda_s^2} \right)$ as
in \cite{Fay68, Kagan88, Baranov92}.

Note that if we change the hierarchy of parameters and make
Coulomb repulsion weaker $W \gg U \gg V$, then in the Born case:
\beq\label{weak_Coulomb_lambdap} \lambda_p \sim
\frac{mVd^2}{4\pi} p_F^3 d^3 \sim \frac{mVd^2}{4\pi} \lambda_s^3
\sim \frac{V}{W}\lambda_s^3 \ll \lambda_s^2
\left(\frac{U+6V}{W}\right)^2, \eeq for $p_F\rightarrow 0$ and
still the overcompensation of the Kohn-Luttinger attraction by
the bare repulsion due to the intersite Coulomb interaction $V$
is impossible. Thus, in the principal approximation in the gas
parameter we restore the results on the possibility of the
$p$-wave superconductivity obtained earlier in the absence of the
intersite repulsion \cite{Fay68, Kagan88, Baranov92}.

\section{2D extended Hubbard model}

In the 2D extended Hubbard model with attractive interaction
($-V<0$) on neighboring sites the vacuum $T$-matrices for the
$s$-wave and $p$-wave channels were obtained in the Refs.
\cite{Kagan94, Kagan02}. After the substitution $-V \rightarrow
V$ they yield for the $s$-wave channel $U_{eff}^s =U+4V$ in the
repulsive case  $U\gg V \gg W$: \beq\label{Ts2D}
\frac{mT_s(\tilde{E})}{4\pi} \approx \frac{(U+4V)
\left(\frac{md^2}{4\pi} \right)}{1+(U+4V)\left(\frac{md^2}{4\pi}
\right)\ln\left( \frac{W \gamma}{|\tilde{E}|} \right)}, \eeq
where $\gamma \sim 1$ is the numerical coefficient. Again, we
assume that we are in the low energy sector when one can neglect
the second pole of the $T_s$ which corresponds to the antibound
state $E\approx U$ \cite{Hubbard63, Anderson90}. In the Eq.
(\ref{Ts2D}) $W=8t$ for the 2D square lattice, and the energy
$\tilde{E}=E+W$ is measured from the bottom of the band. If
$(U+4V)/W \ll 1$, then \beq \frac{mT_s(\tilde{E})}{4\pi} \approx
\frac{1}{\ln\left( \frac{W \gamma}{|\tilde{E}|} \right)}. \eeq In
the Cooper problem $|\tilde{E}|=2\eps_F$ and with the logarithmic
accuracy we restore the 2D dimensionless gas parameter of
Bloom \cite{Bloom75}: \beq f_s \approx \frac{1}{\ln \left(
\frac{1}{nd^2}\right)}, \eeq where $n=p_F^2/2\pi$ is the electron
density in 2D.

Analogously, in the $p$-wave channel $U_{eff}^p = 2 V
\vec{p}\vec{p}'d^2$, and the $p$-wave $T$-matrix reads:  \beq
\frac{mT_p}{4\pi} = \frac{2 mA_p}{4\pi} \vec{p}\vec{p}'d^2 =
\frac{2 mA_p}{4\pi} pp'd^2 \cos\phi, \eeq

where $\phi = \widehat{\vec{p}\vec{p}'}$, and \beq
\frac{mA_p}{4\pi} = \frac{mVd^2}{(1+V/V_{cp})8\pi}. \eeq
Correspondingly \cite{Kagan02}, \beq V_{cp}=11.2 t \approx 1.4 W.
\eeq At $V\gg V_{cp}$, the dimensionless $p$-wave scattering
length in 2D reads \beq
\frac{mA_p}{4\pi}=\frac{md^2V_{cp}}{8\pi}, \eeq and, accordingly,
the dimensionless $p$-wave gas parameter is \beq f_p \sim
\frac{2mV_{cp}d^2}{8\pi} p_F^2 d^2 \sim p_F^2 d^2. \eeq
Thus $f_p
\sim p_F^2 d^2$ again in agreement with general
quantum-mechanical results \cite{Landau77} for slow ($p_F d<1$)
particles in vacuum in the 2D case.

\section{The Cooper problem in 2D at low electron density and in the presence of intersite Coulomb repulsion}

If we restrict ourselves to a very low electron density $n_{el}
d^2 \ll 1$ and quadratic spectrum $\eps(p)-\mu = (p^2 -
p_F^2)/2m$, then in the second order of the $s$-wave gas
parameter the irreducible vertex for the Cooper channel reads:
\beq \tilde{\Gamma} = f_s^2 \Pi (\vec{p}+\vec{k}) \eeq However,
the specific form of the polarization operator on quadratic
spectrum in 2D \cite{Afanasiev62} for $\vec{q}=\vec{p}+\vec{k}$
\beq \Pi(q)=1-\mbox{Re} \sqrt{1-\frac{4p_F^2}{q^2}} \eeq makes
the large Kohn's anomaly ineffective for the SC problem
\cite{Baranov92, Chubukov93}. Indeed, in the SC problem $q\le
2p_F$, $\mbox{Re} \sqrt{1-\frac{4p_F^2}{q^2}}=0$, and thus
$\Pi(q)=1$. Hence, the polarization operator does not depend on
$q$, and correspondingly it does not contain harmonics with
$l\neq 0$ (or more precisely, with the magnetic quantum number
$m\neq 0$). Thus  $\Pi_{m=1}=0$, and SC arises only in the third
order of $f_s$ for quadratic spectrum (or in the second order of
$f_s$ if we take into account corrections $\left(p_x^4 + p_y^4
\right)d^2/m$ which differ the exact spectrum on the square
lattice $\eps(p)=-2t(\cos p_x d+\cos p_y d)\approx -W/2 +p^2/m
-\left(p_x^4 + p_y^4 \right)d^2/24m$ from the quadratic one
$\eps(p)=-W/2 +p^2/m$, see Ref. \cite{Baranov92-2}). At very low
density $n_{el}\rightarrow 0$ the third order terms in the
quadratic spectrum from three irreducible diagrams in the Fig.
\ref{fig: 3_order} dominate over the quartic corrections to the
spectrum.

Chubukov \cite{Chubukov93} found  the leading contribution to
$\tilde{\Gamma}_{m=1}$ from the first skeleton  diagram in which
the Cooper loop is inserted into the polarization loop (it is
important that this diagram is still irreducible with respect to
Cooper channel). Moreover, the character of the large 2D Kohn's
anomaly in this diagram changes and it becomes
$\mbox{Re}\sqrt{2p_F-q}$. Thus, the Kohn's anomaly becomes
effective for SC in the third order. As a result he has obtained
$\tilde{\Gamma}_{m=1} = -4.1 f_s^3$ in the Ref.
\cite{Chubukov93}. In the Ref. \cite{Efremov00-1}, all three
irreducible skeleton diagrams on Fig. \ref{fig: 3_order} were
calculated numerically on equal ground. As a result, the exact
vertex \beq \tilde{\Gamma}_{m=1} = -6.1 f_s^3 \eeq is even a
little bit more attractive. The details of this calculation will
be published in a separate article.

Thus, the total $\tilde{\Gamma}_{m=1}$ at $n_{el}\rightarrow 0$
reads: \beq \tilde{\Gamma}_{m=1} = -6.1 f_s^3 + \alpha p_F^2 d^2
+ o(f_s^4), \eeq where $\alpha \sim 1$ is a numerical
coefficient.

Of course, keeping in mind  that $f_s \sim {1}/\ln
{\left({1}/{nd^2}\right)} \sim {1}/\ln {\left({1}/{p_F^2
d^2}\right)}$, we see that $f_s^3 \gg p_F^2 d^2$ at $p_F d \ll
1$. Thus, $\tilde{\Gamma}_{m=1} \approx -6.1 f_s^3$ just like in
the case $V=0$.

We can see again that in the strong-coupling limit  $U \gg V \gg
W$ of the extended Hubbard model on the square lattice and at low
electron density $p_F d \ll 1$ an inclusion of Coulomb repulsion
does not change the main exponent for the $p$-wave critical
temperature \beq T_{c1}\sim\eps_F \exp\left(-\frac{1}{6.1 f_s^3}
\right). \eeq Thus in the principal order in the gas parameter we
again restore the results on the $p$-wave superconductivity
obtained earlier \cite{Chubukov93, Efremov00-1} in the absence of
the intersite Coulomb repulsion.

\section{Discussions: the case of larger densities}

\begin{figure}[t]
      \centerline{\includegraphics[width=8 cm]{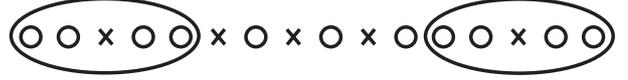}}
    \caption{Fig. 4. Phase separation at the density $n_{el}\leq 1/2$ into metallic droplets in charge-ordered matrix.}
    \label{fig:_phase_separation}
\end{figure}

If we increase the density in the 2D case, we should remember
that at $U \gg V \gg W$ the homogeneous metallic state stretches
only up to the density $n_{el}=\frac{1}{2} -\delta_c$, where in
2D $\delta_c\sim (W/V)^{1/2}$  (see Ref. \cite{Kagan01}). At
$n_{el}>\frac{1}{2} -\delta_c$ the system undergoes a phase
transition into phase-separated state with metallic clusters
inside charge-ordered checkerboard matrix (see Fig.
\ref{fig:_phase_separation}).

Note that at $n_{el}=1/2$  (quarter-filled band) we have Verwey
localization (charge ordering) due to the condition $V\gg W $.
Thus, we cannot extend our calculations for $T_c$ in homogeneous
case to densities larger than $n_{el}=1/2$. However, it is
interesting to construct the SC phase diagram of the extended
Hubbard model with the Coulomb repulsion on neighboring sites at
the intermediate density $n_{el} \le 1/2$, and to find the
regions that correspond to the $p$-wave, $d_{xy}$, and $d_{x^2 -
y^2}$  pairings \cite{Baranov92, Baranov92-2, Kagan94h}.

Another interesting question would be to add to the model an
infinite set of Coulomb repulsion terms with the amplitude
decreasing with the distance between the sites: $V_2 n_i n_{i+2}$
on next-to-nearest sites with $V_2<V$, $V_3 n_i n_{i+3}$ on
next-to-next-to-nearest sites with $V_3 < V_2 < V$  etc. and to
build a bridge between the extended Hubbard model and the jellium
model for screened Coulomb interaction considered in Ref.
\cite{Alexandrov11}.

We think, however, that at least at very low electron density
$n_{el}\rightarrow 0$ our results on the $p$-wave critical
temperature will be stable in the main order of the gas parameter
$p_F d \ll 1$ in 3D and $1/\ln(1/p_F^2d^2)$ in 2D.

\section{Conclusion}

We considered the extended Hubbard model with Coulomb repulsion
on the neighboring sites in the most repulsive (and thus the most
unfavorable for effective attraction and SC) strong-coupling case
$U\gg V \gg W$. In the limit of small electron density $p_F d \ll
1$  we found that the contribution from the intersite Coulomb
repulsion $V$ to the irreducible bare vertex
$\tilde{\Gamma}_{l=1}$  in the $p$-wave channel  is proportional
to $(p_F d)^3$  in 3D and to $(p_F d)^2$ in 2D in agreement with
general quantum-mechanical results for slow particles  in vacuum.

Thus both in 3D and 2D these repulsive terms cannot
overcompensate attractive contributions which are proportional to
$(p_F d)^2$ in 3D and to $1/\ln^3(1/p_F^2d^2)$  in 2D. Note that
the attractive contributions appear only in the presence of
fermion background ($\eps_F \neq 0$). Thus the results of  Refs.
\cite{Baranov92, Chubukov93, Efremov00-1, Efremov00-2} on the
$p$-wave SC of Kohn-Luttinger type \cite{Alexandrov11} both in 3D
and 2D repulsive-$U$ Hubbard model at low electron density and
strong coupling $U\gg W$  are robust against the addition of even
strong Coulomb repulsion on neighboring sites $V\gg W$ in the
extended lattice models. Hence we can see that the $p$-wave
superconductivity exists in purely repulsive models without
electron-phonon interaction.

Note that  we can strongly increase the $p$-wave critical
temperature already at low density in a spin-polarized case
\cite{Kagan89} or in the two-band situation \cite{Kagan91} and
thus reach the realistic values of $T_c$ (of the order of $1 - 5$
K  especially in the 2D or in layered systems \cite{Kagan11}).
The p-wave pairing is realized or can be expected in superfluid
$^{3}$He and ultracold Fermi-gasses, heavy fermion compounds and
Sr$_2$RuO$_4$, semimetals and superlatices, layered
dichalcogenides and organic superconductors
\cite{Vollhardt&Volovik, Kagan96, Maeno01, Murase86}.

\section{Acknowledgements}

We acknowledge useful discussions with A.V. Chubukov, A.S.
Alexandrov, I.A. Fomin, K.I. Kugel and V.V. Kabanov. This work
was supported by RFBR grants 11-02-00708 and 11-02-00741. M.S.M
acknowledges the support by the Department of Energy under Award
Number DE-FG02-08ER64623 (Hofstra University Center for Condensed
Matter).

\end{document}